\documentstyle[prb,preprint,aps]{revtex}
\begin{document}
\draft
\title{{\it Ab initio} Calculations of Multilayer Relaxations of Stepped Cu Surfaces}
\author{Rolf Heid $^1$, Klaus Peter Bohnen $^1$, Abdelkader Kara $^2$,
 and Talat S. Rahman $^{2,3,*}$}
\address{$^1$ Forschungszentrum Karlsruhe, Institut fuer Festkoerperphysik, 76021 Karlruhe, Germany}
\address{$^2$ Department of Physics, Cardwell Hall, Kansas State University,
Manhattan, KS 66506, USA} 
\address{$^3$ Fritz Haber Institut der Max Planck Gesellschaft, 4-6 Faradayweg, 14195 Berlin, Germany}
\address{$^*$Corresponding author: e-mail: rahman@phys.ksu.edu,
 FAX: 785 532 6806}
\date{\today}
\maketitle
\begin{abstract}
We present trends in the multilayer relaxations of
several vicinals of Cu(100) and Cu(111) of varying terrace widths and geometry.
The electronic structure calculations are based on 
density functional theory in the local density approximation with
norm-conserving, non-local pseudopotentials in the
mixed basis representation.  While relaxations continue for several
layers, the major effect concentrates near the step and corner atoms.
On all surfaces the step atoms 
contract inwards, in agreement with experimental findings.
Additionally, the corner atoms 
move outwards and the atoms in the adjacent chain undergo large inward 
relaxation.  
Correspondingly, the largest contraction (4\%) is in the bond length 
between the step atom and its bulk nearest neighbor (BNN), while that
between the corner atom and BNN is somewhat enlarged. 
The surface atoms also display changes in registry of upto 1.5\%.  
Our results are in general in good agreement with LEED data including
the controversial case of Cu(511).  Subtle differences are found with 
results obtained from semi-empirical potentials.
\end{abstract}

PACS\# 61.50.Ah, 68.35.Bs, 68.47.De

\section{\bf INTRODUCTION}

Structural properties of regularly stepped metal 
surfaces have been the focus of a broad
range of theoretical and experimental studies because of the eminent role
they play in technologically important phenomena such as thin film growth,
epitaxial layer formation, nanostructuring of material, and catalysis
 \cite{wan91}.  
According to crystallographic notation, these
surfaces are denoted by high Miller indices and are 
called vicinals of their low Miller index counterparts (flat surfaces). 
The presence of arrays of atomic steps separated by flat terraces
creates regions of differing local coordination and  
makes the microscopic structure of a 
vicinal surface distinct from that of a flat surface.  According to
Smoluchowski's idea of charge smoothing \cite{smo41}, for example,
electronic charge densities are expected to
rearrange in the vicinity of the steps, thereby causing the ion cores to 
relax to new configurations. 
The modified electronic structure may also be expected to impact the reactivity
and the nature of the force fields in the region around the steps. 
Knowledge 
of atomic relaxations in the equilibrium positions near the step and kink
sites is thus a step towards understanding the novel vibrational and
electronic properties of vicinal surfaces.

Fortunately with advances in atomic scale experimental techniques, there 
has been a surge
in investigations of the structure of vicinal surfaces in  recent years. 
The majority of the experimental data have undoubtedly come 
\cite{struc_exp,Hove}
from the low-energy electron diffraction (LEED) technique which is now capable
of detecting changes even for interlayer spacings smaller than
1$\AA$\cite{jon99}.  For some surfaces the X-ray scattering technique has 
provided much needed complementary structural data \cite{wal99}.
An impressive number of theoretical calculations of multilayer
 relaxations \cite{vic_rel,nel98,tia93,dur97,skl98} have also
helped in bringing several issues 
 related to the characteristics of vicinal surfaces to the forefront.   
Of particular 
interest here are experimental and theoretical studies of a set of
vicinals of Cu(100) and Cu(111)  
which have addressed the question of the impact of local coordination
on the structural and dynamical properties of the surface.
  In an earlier paper \cite{dur97}, 
 a comparative study of the local structural and vibrational 
properties of Cu(211), Cu(511), and Cu(331) was performed using empirical 
potentials from the embedded atom method (EAM) \cite{foi86}.  This study found 
that the first two surfaces displayed similar local  characteristics, 
while the third surface was somewhat different.
 An explanation provided for this behavior was the similarity in the
 local environment of the (211) and (511) surfaces of fcc metals (a
combination of (100) and (111), terrace geometry, and step face), and its
consequent difference from that of the (331) surface( a (111) terrace 
 geometry and a (111)-microfacetted step face).
  The issue of the impact of the local geometry was 
further raised in a joint theoretical and experimental study of the 
vibrational dynamics of Cu(211) and Cu(511) together with
those of the kinked surface Cu(532) \cite{kar00}.  Experimental data from 
Electron Energy Loss Spectroscopy (EELS) found modes above the bulk band
on Cu(211) but not on Cu(511) (or on  Cu(17,1,1) which has the
 same step geometry as Cu(511)), but theoretical calculations
based on EAM potentials predicted modes (slightly)
 above the bulk phonon spectrum 
for each of these surfaces.  While the similarity between the calculated 
structural relaxation patterns of Cu(211) and Cu(511) argues in favor of a
  similarity in the local vibrational dynamics of these two surfaces, 
 the disagreement between the experimental and the theoretical results for the 
high frequency modes on Cu(511) (and Cu(17,1,1)) remains unreconciled.  
For Cu(211) agreement of the EAM
based results with available structural data from LEED  \cite{sey99}
and with {\it ab initio} calculations for both the structure and the
dynamics \cite{wei98}  
provides considerable confidence in its predicted properties.
  The case of Cu(511) is not as simple because 
of lack of calculations based on potentials more accurate than EAM, and because 
of conflicting conclusions from the analysis of experimental data from 
LEED \cite{alb01} and
X-ray scattering measurements \cite{wal99}.  The most striking difference 
in these two sets of data is the relaxation pattern for the second 
layer which is inwards in LEED and outwards in the X-ray data. 
 The oscillatory pattern found in the X-ray data is
also in disagreement with the conclusion from a series of previous 
experimental and theoretical findings on stepped surfaces.  
Based on these studies \cite{nel98,tia93,dur97,skl98,jon00}, 
there is a definite 
symmetry in the relaxation patterns of stepped surfaces.  All terrace atoms,
save for the corner one, display inward relaxations.
The EAM based calculations \cite{tia93} further predict this
oscillatory relaxation pattern to continue into the bulk with a damping
in the amplitude \cite{all88}.
Thus the expected relaxation pattern for the (211), (511), (331) surfaces,
each with 3-atom wide terraces, would be  (- - + - - +...), although
questions have been raised whether Cu(331) follows this rule \cite{dur97}.
Similarly, the patterns for (711) and (911) with,
 respectively, 4 and 5 atoms on the terrace, would 
 be predicted to be (- - - + - - - +...) and (- - - - + - - - - +...).  
The LEED data on the first three surfaces follow these predicted trend in
 relaxations, atleast for the top 3 layers.  The very recent LEED data
\cite{wal01} for Cu(711) also displays the pattern (- - - +) for the 
top layers, in good agreement with EAM based predictions.
However, a small discrepancy in the sign of the relaxation is found 
for both Cu(711) and Cu(511), for a particular layer separation (d$_{56}$ for
Cu(511) and d$_{78}$ for Cu(711)) \cite{wal01}, beyond that expected from 
the error bars.  Arguably the actual numbers involved in these comparisons
are small, but the systematic nature of the
 discrepancies and the fact that it negates the prediction of a periodicity
 in the oscillatory relaxation pattern \cite{tia93,all88}, raise interesting
 questions about the complexities of the atomic
 displacements in these systems.
Given the  above uncertainties arising from experimental
observations,  it is opportune to carry out more accurate
calculations of these relaxation patterns using techniques which are  
capable of revealing the accompanying changes in the surface electronic 
structure.
It is with this goal in mind that we have carried out {\it ab initio} 
electronic structure calculations of the surface geometry and interlayer
spacing for a set of vicinals of Cu(100) and Cu(111).
 In addition to Cu(211), Cu(331) and Cu(511) which are   included 
 to address the question of the influence of the local geometry
on the structure,  we have extended the investigation to Cu(711)
and Cu(911)  to examine the influence of increasing terrace width
of the relaxation pattern. Of course, for all surfaces comparison
 with available experimental data is of prime concern.

The rest of this report is organized as follows.  In Section II, the system 
geometries are presented together with some computational details.  Section III
contains the results and their discussion.  Concluding
remarks are presented in Section IV.

\section{\bf SURFACE GEOMETRIES}

Vicinal surfaces can easily be  constructed by cutting the crystal
at an angle slightly away from the lower-index crystal planes (i.e. (100),
 (111), (110)).   For reasons discussed above, 
we are interested here in the vicinals
of the (100) and (111) surfaces of fcc metals 
  of which the most tightly packed
steps are along the $<110>$ direction.  In the case of the (111) surface,
 however, the $<110>$ direction is not parallel to any plane of symmetry
and  there are two different ways of generating monoatomic
stepped surfaces. In one type of such vicinals, the step edge has
 a (100)-microfacet, while the other has the (111)-microfacet
(these are the so called A and B types, respectively).
  In the standard nomenclature, the vicinals
of fcc(111) surface with monoatomic steps and (100) step edges are
denoted by $(n,n,n+2)$, while those with (111) step edges are  labeled as
$(n,n,n-2)$, where $n$ is the number of atoms on the terrace.  The B-type 
vicinal Cu(331) considered here is named
accordingly, while the A-type vicinal Cu(211) seems to be a misnomer.
Similarly, the vicinals of fcc(100) surface consisting of
monoatomic step edges with (111) microfacet are
labeled $(2n-1,1,1)$.  The Cu(111) vicinals considered here are  
created by cutting the crystal at an angle of 19.5$^o$
and 22$^o$ away from the (111) plane towards the [$2\bar{1}\bar{1}$]
and [$\bar{2}11$] direction to produce the (211) and (331)
 surfaces, respectively, whereas the three
 vicinals of (100), (511),
 (711), and (911), are constructed by slicing the
crystal at angles of 15.8$^o$, 11.4$^o$ and 8.9$^o$, respectively, 
 from the (100) plane towards the
[011] direction. To facilitate the discussion we have also labeled the atoms
 that play the dominant role in our calculation \cite{dur97}.
For the three surfaces consisting of three chains of atoms 
on the terrace   we label them as corner-chain (CC), terrace-chain (TC1),
and step-chain (SC).  The chain just underneath
the step-chain is called a bulk nearest neighbor chain (BNN).  
The other two surfaces,  Cu(711) and Cu(911), 
contain, respectively, one and two extra chains of terrace atoms, labeled
 accordingly as (TC2) and (TC3).
We have taken the $x$ and $y$ axes to lie in the surface plane, the
$x$-axis being perpendicular to the step and the $y$-axis along
the step, and the $z$-axis is along the surface normal.
 In Fig. 1, we display a side view of the (511) surface with the
 appropriate labeling of the atoms and interlayer spacing.

\section{\bf SOME DETAILS OF THEORETICAL CALCULATIONS}

The {\it ab initio} electronic structure
 calculations are performed within a pseudopotential
approach to density-functional theory in the local
 density approximation \cite{hoh64}, numerical implementation
 of the technique is based on a computer code developed by B. Meyer 
 {\it et al.}  \cite{meyer} The local-density approximation
 is applied using the Hedin-Lundqvist form of the exchange-correlation 
functional \cite{hed71}. 
A norm conserving pseudopotential for Cu constructed according
to a scheme proposed by Hamann-Schlueter-Chiang \cite{ham79} 
 has been used which has already been successfully employed for 
 calculations of the structure and the
phonons of low index surfaces of Cu \cite{rod93}.
 A mixed basis set is applied to 
represent the valence states consisting of five d-type local functions at
each Cu site, smoothly cut off at a radius of 2.3 a.u.,
 and of plane waves with
 kinetic energy of 11 Ry. The Brillouin-zone(BZ) integration was
carried out using the special point sampling technique
 \cite{mon76} together with
a Gaussian broadening of the energy levels of 0.2 eV. For simulating 
surfaces we used the supercell approach with cells containing 21 to 35 atoms
 (1 atom per layer), depending on the surface orientation.
The z-dimension of all cells was 47.7155 a.u.
The distance between the top and the bottom layer of the slabs were thus
31.2761 a.u. for Cu(331), 27.8277 a.u. for Cu(211), 26.2340 a.u. for
Cu(511), 25.7715 a.u. for Cu(711), and 25.4390 a.u. for Cu(911).
Structure optimization
was carried out until forces on all atoms were smaller than $10^{-3}$
Ry/a.u., which is two orders of magnitude smaller than 
 the forces present on  the unrelaxed
surfaces. With increasing terrace width, the calculations become
increasingly tedious since the reduction in interlayer spacing makes it
more difficult to achieve geometries converged to 1\% of the interlayer 
spacing.  We also find that results for terraces with (100) geometry are more
sensitive to the number of k-points sampled,
 as compared to those with (111) geometry.
For the latter case 30 points in the BZ are sufficient for the
 determination of the equilibrium structure, while for surfaces
with (100) terraces at least twice as many points are needed to get
converged results.

\section{\bf RESULTS AND DISCUSSIONS}

Our results for the multilayer relaxations of Cu(211), Cu(331), Cu(511),
Cu(711) and Cu(911) are summarized in Table I.  As in previous theoretical
studies \cite{tia93,dur97,skl98,kar00} of relaxations on stepped Cu surfaces,
we find changes in interlayer separations, from bulk terminated 
configurations, to persist on all surfaces for a large number of layers.
Of course, the concept of layers is different here from that on flat
surfaces (see Fig. 1). 
  The first {\it n }  layers, for example, are all
exposed to the vacuum, where {\it n } is the number of atoms on the terrace.
Correspondingly, the interlayer separations are small and even large
percentage changes in interlayer separations correspond to small 
numbers in distances.  It is important to bear this point in mind
when comparing the results for a particular surface either with those
for flat surfaces or with the results from other methods.

A common feature of all surfaces examined in Table I is that all terrace
atoms except for the ones in the corner chain (CC) undergo significantly
large inwards relaxations.  The corner atoms are always found to relax
outwards.  Additionally, the atoms in the terrace adjacent to CC exhibit
comparatively large inward relaxation whose magnitude maybe larger than
that of the step atoms (SC).  For example, for Cu(511), Cu(711), and 
Cu(911) changes, respectively, in $d_{23}$ (involving TC1), 
$d_{34}$ (involving TC2), and $d_{45}$ (involving TC3), are
larger than that of $d_{12}$.  Thus, in keeping with Schmolkowski's \cite{smo41}
ideas of charge smooting, the maximum relative change in interlayer
separation is focussed around the corner atoms.  This is particularly
true for the vicinals of Cu(100).  The situation with the more closepacked
surface Cu(331) is somewhat different, As seen in Table I
 the outward relaxation of the corner atom and the inward
 relaxation of the preceeding atom on the terrace on Cu(331)
 are less than half of that for similar atoms on the other surfaces
 considered here.  Incidently, this conclusion is in good agreement 
 with results from previous
studies which were based on semi-empirical potentials \cite{tia93,dur97}. 

 There is, however,
a disconcerting difference in the results obtained here from {\it ab initio}
electronic structure calculations and those from 
 semi-empirical potentials.  An intriguing result for multilayer
relaxations of the vicinals of Cu(100) obtained with EAM potentials in
Ref. \cite{tia93} was that the pattern of inward and outward relaxations
continued well into the bulk with an expontially decreasing amplitude.
Thus for Cu(711), the relaxation pattern predicted by EAM was 
(- - - +,- - - +,- - - +,...) with an eventual dampimg of the relaxations. 
The pattern for Cu(711) from Table I is instead (- - - +,- - + +,- - + +).
That is our present calculations do not predict a periodically
  oscillatory relaxation
pattern with a decaying amplitude as we move into the bulk.  
As we shall see, this particular feature is more in agreement with
experimental data and help remove the slight discrepancy between
experiment and theory presented by the EAM result pointed by Walter et al.
\cite{wal01}.
 Again, it should be recalled that the numbers involved are very small
 and within the limits of accuracy of {\it ab initio} calculations.
 In particular, the small numbers for the relaxations of the inner 
 layers of Cu(911) have to be taken with caution as our convergence criteria 
 for this surface were not as good as that for the others because of 
 the demands 
 on computational resources imposed by a system as large as this one.

Unlike flat surfaces, vicinal surfaces relax in both $x$ and $z$ directions,
since the existence of steps at the surface leads to broken symmetry
in both of these directions.  While relaxations along the $z$-direction yield
characteristic interlayer separations that we have discussed above, those 
along the $x$-direction provide new registries of atoms, as compared to those 
in the bulk.  Our calculated percentage intralayer registries for five
surfaces are summarized in Table II.  As in the observations from EAM 
calculations, the changes in the registries of the atoms are small.  It is
thus not useful to make a one-to-one comparison with results from semiempirical
calculations.  However, the changes in registries of the atoms are not
inconsequential since they affect the changes in the bond lengths between
the atoms in these regions of low coordinations.  In Table III, we tabulate
our results for the total changes in the distances between the step atoms 
and their nearest neighbors.
For comparison we have included in parenthesis the results obtained earlier
for the same quantities with EAM potentials \cite{dur97,tia93}.  The largest
changes in the bond lengths (from unrelaxed configurations) are for those
between the step atoms and their bulk nearest neighbour (BNN) which lies
right below them.  The bonds between CC and BNN show small enlargement,
while all other bonds in Table III are found to undergo shortening.
 In Figs. 2 and 3, we have drawn the actual displacements of the atoms
 on the five surfaces obtained from our calculations. While the size of 
the arrows are exagerated, it is their relative length and direction
 that is of consequence. 
  As already noted by Durukanoglu et al. \cite{dur97}, all atoms in
  low coordinated sites move to enhance their local coordination.
  The complex displacement pattern that emerges is thus the net outcome of
  the competition between the different directions in which the various
  atoms would like to relax to enhance thein own coordination.
 For readers who are interested in the exact positions of the atoms
 on the relaxed surfaces, we have summarized them in Table IV.
 The unusual behavior of Cu(331) terrace atom is more apparent from this
 Table than the earlier one on changes in the bond lengths.  The TC1 
 atom of Cu(331) undergoes the least displacement among its counterparts.
 Its displacement is also smaller than that of TC2 (1.4, 0.0, -2.6)
 on Cu(711), and of both TC2 (0.8, 0.0, -2.0) and TC3 (0.1, 0.0, -2.3)
 on Cu(911).  The coordinates of the displacements above in paranthesis 
 are in the same units as those in Table IV.

We now turn to comparisons of the results obtained here for individual
surfaces with those available from experimental measurements. 
In Table V we show that for Cu(211) the salient features in the
 trends in the relaxation patterns predicted by our calculations
 are observed in the experimental data. Apart from the large inward 
 relaxation of the step atoms, the major change occurs at the corner atom
 and its adjacent terrace atom. Our results are in good
 agreement with previous DFT/LDA calculations \cite{wei98},
 based on the pseudopotential approximation and with results
 from EAM based method. Theoretical calculations using the full potential
 linearized augmented plane wave (FLAPW) method \cite{geng}, however, 
predict much larger relaxation(-28.4\%) of the step atom than any of
the previous theoretical or experimental studies.
  This brings us to the discussion of ionic relaxation on 
 Cu(331) in Table VI  for which also we do not
 get the large relaxation reported in Ref[\cite{geng}]. 
  The present results for Cu(331) are, however, in good agreement with the 
  LEED data \cite{jon00}.  The discrepancy with the results
 from LEED for $d_{23}$ should not be taken too seriously, given
 an error bar of 4\% in the analysis of the LEED data.
With respect to EAM based results \cite{dur97}, we find a noticeable 
 difference for d$_{34}$, for which the present results agree
 better with the LEED data and also preserve the predicted relaxation pattern
 (- - +) for the terrace atoms.  This trend is inkeeping with what was
 reported in calculations on  Al(331) \cite{nel98}.
  In trying to reconcile our results with those of
 Geng et al. \cite{geng}, we note that the latter predict an outward 
 displacement of the TC1 atomic
  chain for both Cu(211) and Cu(331), while we find this not to be the case.
 As already mentioned, while the changes in the bond lengths of the terrace
 atoms of Cu(331) are no different from those of the other
 surfaces, the displacement of TC1 is strikingly smaller than that
 of the TC's on other surfaces.

 The case of multilayer relaxations for Cu(511) is interesting because of 
 the differences in the published data from LEED \cite{alb01} and 
 X-ray measurements \cite{wal99}.  These are displayed in
 Table VII. Except for the displacement of the
 step atoms, for which all results point to a large inward relaxation,
 the results from X-ray scattering measurments are in disagreement
 with present results and with those from LEED, as well as, from EAM
 calculations. We do not understand the reasons for this disagreement
 but for the notion that X-ray measurements may be very sensitive
 to the quality of the crystal surface. It should be noted that
 the differences with the X-ray results
  are both qualitative and quantitative,
 beyond the established error bars in the experiments and calculations.
 Because of the controversy in the experimentally determined 
 multilayer relaxations of Cu(511), we have carried out an extensive
 analysis of the dependence of the theoretical results on the
 approximations necessary to produce computational feasibility: 
 choice of pseudopotentials, maximum kinetic
 energy of the plane-waves (E$_{cut}$), the number of layers
 in the supercell, and the number of points used to sample the surface
 Brillouin zone. As for the dependence of the results on the type
 of pseudopotentials and E$_{cut}$, we have carried out calculations 
 with three different  pseudopotentials (on three sets of codes) to
 find that the results which lie within 3\% of each other.
  We also find our choice of supercell size
 to be adequate. There is, however, a strong dependence of the 
 results on the number of BZ points sampled. For the case of Cu(511) this
 dependence is illustrated in Table VIII. Calculations performed with
few points could give erroneous relaxations as signified 
 by the case of d$_{23}$ in Table VIII. An inward relaxation of 1.8\%
 is found with 4 points, while the converged result is 10.7\%.
 Convergence in the calculated relaxation is reached once the
 number of points is increased to 24 and beyond. Thus, when comparing
 results from {\it ab initio} calculations, one has to keep
 these technical points in mind. Unless sufficient checks are made 
 for convergence
 in the reported values, quantities like equilibrium positions of surface atoms
 may differ in different calculations and lead to disagreement in the 
 calculated relaxations.  It would be worthwhile to clarify whether
the differences between our results and those from the FLAPW method
 for Cu(211) and Cu(331) could be attributed to k-points sampling. 

 Finally, we come to the comparison of our results for Cu(711) with 
 those from experiments ( we are not aware of any data on Cu(911), so far).
 The LEED data \cite{wal01} for this surface has been very carefully 
 analyzed and compared to existing calculations.
 Table IX shows that the {\it ab initio} results obtained
 here are in excellent agreement with the data, and that the small
 differences with the EAM results that the authors\cite{wal01} had noted,
 is removed by the present calculations.
 As in the case of Cu(511), the largest percentage change in the
 interlayer spacing is not for d$_{12}$. In this case it is for d$_{34}$
 which separates CC from TC2. As before, there is outward relaxation of the 
 spacing between CC and BNN. The fact that relaxations near CC persist 
 on being strong even as the terrace width increases, is interesting in itself.
 This particular argument has not been made in any previous theoretical result.
 Our calculated values for Cu(911) further support this argument as the largest
 percentage change is found for d$_{45}$, the interlayer spacing between
 CC and TC3 (in this case). While these results are intriguing the main
 outcome of relaxations that ensue when a surface is created is in the actual
 displacements of the atoms from their bulk terminated positions to the
 new equilibrium positions. As already stated, these values
 are summarized in Table V and the related patterns presented
 in Figs. 2 and 3. Obviously, for stepped surfaces
 there is a complex rearrangement of most terrace atoms. Our
 calculations show that despite this complexity, all terrace atoms except
 for CC move inwards.

\section{CONCLUSION}

In summary, we have performed a
 comparative study of multilayer and atomic relaxations
of five stepped Cu surfaces which are vicinals of Cu(100) and Cu(111) using
{\it ab initio} electronic structure calculations based on density functional 
theory and non-local,
 norm-conserving pseudopotentials.  The set of three of these surfaces:
Cu(211), Cu(331), and Cu(511), provides a comparison of structural changes from
bulk termination,
 for vicinals of similar terrace widths but differing local geometry.
The other set consisting of Cu(511), Cu(711) and Cu(911) offers a comparative 
study of relaxation patterns
 with changing terrace width.  In each case we find the 
relaxation of the step atoms to 
be pronounced inwards and that of the corner atom to be
 outwards.  The other terrace
atoms and their nearest neighbors
 also undergo relaxations following a complex displacement
 pattern. Subsequently, the bond lengths between all the
 surface atoms and their nearest neighbors change from
 the bulk terminated values while the bond length between CC
 and BNN atoms experiences an elongation (about 1\%) all other surface
 length shrink anywhere from 1\% to 4\%. Most of our
 findings are in agreement with previous calculations which
 were based on semiempirical model potentials except
 that we do not find the pattern of inward relaxations
 of SC, TC1, TC2 etc followed by outward relaxation of CC atoms
 to continue into inner layers. We also find that the percentage
 contraction of the spacing between the TC and CC atoms is generally
 larger than that between SC and the TC atoms. While the actual
 magnitudes of the changes in the spacing considered here are small,
  there is a systematic trend in the relaxation pattern which
points to significant rearrangements in the electronic
 charge densities near the SC and CC atoms.
 By and large our results are in good agreement with available
 structural data on these surfaces, except for the case of Cu(511)
 for which we favor the LEED results over those from X-ray scattering 
meaurements. We believe our results will help settle the issues
 that have emerged on this particular surface.
 Our systematic examination of five surfaces, also helps address the question wether
 the relaxations on Cu(331) are anomalous. The only striking difference
 between this surface and the others is in the relaxation of TC1
 which is very small. Otherwise the relaxation pattern and the changes
 in bond lengths are similar to those on the other surfaces. 

The main message from these observations is that the important quantity 
to examine is the displacement pattern of the surface atoms as they 
relax to their
 equilibrium positions from their bulk terminated configurations.
 The deeper question, of course, is the nature of the accompanying 
 changes in the surface electronic structure. 
 It will be interesting to examine the characteristics of the local 
 electronic densities of states in the different regions of low symmetry 
 that are present naturally on the stepped surfaces considered
 here. We leave this as an exercise for the future.

\section{ACKNOWLEDGMENTS}
The work of TSR and AK was supported in part by the US National Science
Foundation, Grant CHE-9812397 and by the Basic Energy 
Research Division, Department of Energy, Grant DE-FG03-97ER45650.  
TSR also acknowledges 
the Alexander von Humboldt Foundation for the award of a Forschungspreis
 and thanks her colleagues at the Fritz Haber Institut, 
Berlin and at the Forschungszentrum, Karlsruhe for their warm  hospitality.


%
\begin{figure}
\caption{Sideview of fcc(511) surface showing the interlayer separations,
and the labeling of the atoms and the layers.}
\label{Fig. 1}
\end{figure}

\begin{figure}
\caption{Schematic representation of displacements of atoms during the 
relaxation process for Cu(511), Cu(711), and Cu(911).}
\label{Fig. 2}
\end{figure}

\begin{figure}
\caption{Schematic representation of displacements of atoms during the 
relaxation process for Cu(331) and Cu(211).}
\label{Fig. 3}
\end{figure}

\begin{table}
\caption{ Calculated changes in interlayer separations as percentage
 of the ideal separation $d_{b}$.}
\label{tab:relax}
\begin{tabular}{cccccc}
\multicolumn{1}{l}{Layer}&
\multicolumn{1}{c}{Cu(211)}&
\multicolumn{1}{c}{Cu(331)}&
\multicolumn{1}{c}{Cu(511)}&
\multicolumn{1}{c}{Cu(711)}&
\multicolumn{1}{c}{Cu(911)} \\
\hline
$d_{b}$&$0.736  \AA$& $0.828  \AA$& $0.694 \AA$& $0.505 \AA$&$0.443 \AA$\\
$d_{12}$&$-12.2\%$& $-12.7\%$& $-9.3\%$& $-7.3\%$& $-11.2\%$\\
$d_{23}$&$-9.5\%$& $-3.3\%$& $-10.7\%$& $-1.5\%$& $-2.2\%$\\
$d_{34}$&$+8.7\%$& $+4.5\%$& $+7.2\%$& $-14.8\%$& $+0.6\%$\\
$d_{45}$&$-2.1\%$& $-2.0\%$& $-2.9\%$& $+8.0\%$& $-13.9\%$\\
$d_{56}$&$-1.6\%$& $+0.1\%$& $+1.1\%$& $-1.0\%$& $+5.4\%$\\
$d_{67}$&$+1.5\%$& $-0.1\%$& $+1.7\%$& $-1.1\%$& $-1.3\%$\\
$d_{78}$&$-0.1\%$& $+0.8\%$& $-1.5\%$& $+1.4\%$& $-4.1\%$\\
$d_{89}$&$-0.3\%$& $-0.6\%$& $+1.6\%$& $+1.7\%$& $+4.5\%$\\
$d_{9,10}$&$+0.7\%$& $+0.9\%$& $-0.5\%$& $-1.5\%$& $+3.0\%$\\
$d_{10,11}$& & & & $-0.4\%$& $-0.5\%$\\
$d_{11,12}$& & & & $+2.0\%$& $-2.5\%$\\
$d_{12,13}$& & & & $+0.3\%$& $+1.2\%$\\
$d_{13,14}$& &  & & & $+1.6\%$\\
$d_{14,15}$& &  & & & $+3.0\%$\\
$d_{15,16}$&& & & & $-2.2\%$\\
\end{tabular}
\end{table}

\begin{table}
\caption{ Calculated changes in the registries as percentage of that for
the ideal surface $r_{b}$. }
\label{tab:registry}
\begin{tabular}{cccccc}
\multicolumn{1}{l}{Registry}&
\multicolumn{1}{c}{Cu(211)}&
\multicolumn{1}{c}{Cu(331)}&
\multicolumn{1}{c}{Cu(511)}&
\multicolumn{1}{c}{Cu(711)}&
\multicolumn{1}{c}{Cu(911)} \\
\hline
$r_{b}$&$2.083  \AA$& $2.048  \AA$& $2.454 \AA$& $2.500 \AA$&$2.519 \AA$\\
$r_{12}$&$-1.22\%$& $-0.10\%$& $-1.17\%$& $-1.37\%$& $+0.99\%$\\
$r_{23}$&$-0.54\%$& $-1.74\%$& $-1.21\%$& $-0.32\%$& $-0.40\%$\\
$r_{34}$&$-0.22\%$& $+1.46\%$& $+0.98\%$& $-0.41\%$& $+0.28\%$\\
$r_{45}$&$+1.50\%$& $+0.62\%$& $+0.25\%$& $+0.77\%$& $-0.60\%$\\
$r_{56}$&$-0.26\%$& $-0.50\%$& $-0.31\%$& $+0.23\%$& $+0.38\%$\\
$r_{67}$&$+0.19\%$& $+0.24\%$& $+0.01\%$& $+0.82\%$& $-0.01\%$\\
$r_{78}$&$0.00\%$& $-0.19\%$& $0.00\%$& $-0.89\%$& $+0.55\%$\\
$r_{89}$&$-0.11\%$& $+0.23\%$& $-0.14\%$& $-0.01\%$& $+0.16\%$\\
$r_{9,10}$&& & & $-0.31\%$& $-0.44\%$\\
$r_{10,11}$&& & & $-0.20\%$& $-0.14\%$\\
$r_{11,12}$&& & & $+0.11\%$& $+0.40\%$\\
$r_{12,13}$&& & & & $-0.23\%$\\
$r_{13,14}$&& & & & $-0.25\%$\\
\end{tabular}
\end{table}

\begin{table}
\caption{Changes (in \%) in bond-lengths between step atom and its
nearest neighbors. The results from EAM are in paranthesis}
\label{tab:bondlength}
\begin{tabular}{ccccc}
\multicolumn{1}{l}{Surface}&
\multicolumn{1}{c}{SC-TC}&
\multicolumn{1}{c}{SC-CC}&
\multicolumn{1}{c}{SC-BNN}&
\multicolumn{1}{c}{CC-BNN} \\
\hline
Cu(211)&-1.78(-1.27)& -2.27(-2.67) & -3.22(-2.10) & +0.61(+0.7)\\
Cu(331)&-1.36(-0.45)& -2.39(-1.86) & -3.66(-3.09) & +1.42(+0.5)\\
Cu(511)&-1.80(-2.29)& -1.42(-0.98) & -3.13(-2.30) & +1.49(+1.5)\\
Cu(711)&-1.59(-2.16)& -1.36(-1.11) & -2.88(-2.26) & +1.06(+1.8)\\
Cu(911)&-1.22(-2.08)& -1.47(-1.06) & -3.06(-2.31) & +0.51(+1.93)\\
\end{tabular}
\end{table}

\begin{table}
\caption{Atomic displacements from bulk terminated
to relaxed positions ($\rm \AA (x10^{-2})$)}
\label{tab:displacement}
\begin{tabular}{ccccc}
\multicolumn{1}{l}{Surface}&
\multicolumn{1}{c}{SC}&
\multicolumn{1}{c}{TC1}&
\multicolumn{1}{c}{CC}&
\multicolumn{1}{c}{BNN} \\
\hline
Cu(211)&(-1.8,0.0,-10.9)&(0.7,0.0,-1.9)&(1.7,0.0,5.1)&(2.3,0.0,-1.4) \\
Cu(331)&(-0.4,0.0,-10.0)&(-0.1,0.0,0.5)&(3.5,0.0,3.2)&(0.6,0.0,-0.8) \\
Cu(511)&(-2.5,0.0,-9.2)&(0.1,0.0,-2.7)&(2.7,0.0,4.7)&(0.2,0.0,-0.3) \\
Cu(711)&(-2.8,0.0,-7.1)&(0.6,0.0,-3.4)&(2.4,0.0,4.8)&(0.5,0.0,0.8) \\
Cu(911)&(-2.7,0.0,-7.4)&(-0.2,0.0,-2.9)&(1.6,0.0,3.2)&(0.7,0.0,1.1) \\
\end{tabular}
\end{table}

\begin{table}
\caption{ Relaxation of Cu(211): experiment and theory}
\label{tab:211rlx}
\begin{tabular}{cccccc}
\multicolumn{1}{l}{Relaxation}&
\multicolumn{1}{c}{This work}&
\multicolumn{1}{c}{FLAPW \cite{geng}}&
\multicolumn{1}{c}{EAM \cite{dur97}}&
\multicolumn{1}{c}{DFT-PW \cite{wei98}}&
\multicolumn{1}{c}{LEED \cite{sey99}} \\
\hline
$d_{12}$&$-12.2\%$& $-28.4\%$& $-10.3\%$& $-14.4\%$& $-14.9\%$\\
$d_{23}$&$-9.5\%$& $-3.0\%$& $-5.41\%$& $-10.7\%$& $-10.8\%$\\
$d_{34}$&$+8.7\%$& $+15.3\%$& $+7.26\%$& $+10.9\%$& $+8.1\%$\\
$d_{45}$&$-2.1\%$& $-6.6\%$& $-5.65\%$& $-3.8\%$& \\
$d_{56}$&$-1.6\%$&  $+0.7\%$& $-1.2\%$& $-2.3\%$& \\
$d_{67}$&$+1.5\%$& $+3.0\%$& $+4.0\%$& $+1.7\%$& \\
$d_{78}$&$-0.1\%$& & $-2.6\%$& $-1.0\%$& \\
$d_{89}$&$-0.3\%$& & $-0.17\%$& $-0.05\%$& \\
$d_{9,10}$&$+0.7\%$& & $+0.0\%$& $+0.0\%$& \\
\end{tabular}
\end{table}

\begin{table}
\caption{ Comparison of multilayer relaxation of Cu(331): experiment and theory}
\label{tab:331rlx}
\begin{tabular}{ccccc}
\multicolumn{1}{l}{Relaxation}&
\multicolumn{1}{c}{This work}&
\multicolumn{1}{c}{FLAPW \cite{geng}}&
\multicolumn{1}{c}{EAM \cite{dur97}}&
\multicolumn{1}{c}{LEED \cite{jon00}} \\
\hline
$d_{12}$&$-12.7\%$& $-22.0\%$& $-10.4\%$& $-13.8\%$\\
$d_{23}$&$-3.3\%$& $+1.6\%$&$+1.7\%$& $+0.4\%$\\
$d_{34}$&$+4.9\%$& $+6.9\%$&$-1.7\%$& $+4.0\%$\\
$d_{45}$&$-2.0\%$& $-2.4\%$&$-0.3\%$& $-4.0\%$\\
$d_{56}$&$+0.1\%$& $-0.6\%$&$-0.3\%$& \\
$d_{67}$&$-0.1\%$& $-0.4\%$&$+0.5\%$& \\
$d_{78}$&$+0.8\%$& &$-0.4\%$& \\
$d_{89}$&$-0.6\%$& & $+0.2\%$& \\
$d_{9,10}$&$+0.9\%$& & $+0.0\%$& \\
\end{tabular}
\end{table}

\begin{table}
\caption{ Relaxation of Cu(511): experiment and theory}
\label{tab:511rlx}
\begin{tabular}{ccccc}
\multicolumn{1}{l}{Relaxation}&
\multicolumn{1}{c}{This work}&
\multicolumn{1}{c}{EAM \cite{dur97}}&
\multicolumn{1}{c}{LEED \cite{alb01}}&
\multicolumn{1}{c}{X-ray \cite{wal99}} \\
\hline
$d_{12}$&$-9.3\%$& $-9.5\%$& $-13.2\%$& $-15.4\%$\\
$d_{23}$&$-10.7\%$& $-7.9\%$& $-6.1\%$& $+8.1\%$\\
$d_{34}$&$+7.2\%$& $+8.8\%$& $+5.2\%$& $-1.1\%$\\
$d_{45}$&$-2.9\%$& $-4.2\%$& $-0.1\%$& $ -10.3\%$ \\
$d_{56}$&$+1.1\%$& $-4.0\%$& $+2.7\%$& $+5.4\%$\\
$d_{67}$&$+1.7\%$& $+3.4\%$& & $-0.7$\\
$d_{78}$&$-1.5\%$& $-1.7\%$& & $-6.9\%$\\
$d_{89}$&$+1.6\%$& $-1.1\%$& & $+3.6\%$\\
$d_{9,10}$&$-0.5\%$& $+0.0\%$& & \\
\end{tabular}
\end{table}

\begin{table}
\caption{ Relaxation of Cu(511):  effect of the number of k-points in the BZ}

\begin{tabular}{ccccc}
\multicolumn{1}{l}{Interlayer}&
\multicolumn{1}{c}{65 k-pts}&
\multicolumn{1}{c}{44 k-pts}&
\multicolumn{1}{c}{24 k-pts}&
\multicolumn{1}{c}{4 k-pts} \\
\hline
$d_{12}$&$-9.3\%$& $-9.9\%$& $-10.6\%$&$-15.1\%$\\
$d_{23}$&$-10.7\%$& $-10.6\%$& $-11.2\%$&$-1.8\%$\\
$d_{34}$&$+7.2\%$& $+7.3\%$& $+7.9\%$&$+5.4\%$\\
$d_{45}$&$-2.9\%$& $-3.4\%$& $-4.1\%$&$-0.1\%$\\
$d_{56}$&$+1.1\%$& $+1.0\%$& $+0.5\%$&$+1.4\%$\\
$d_{67}$&$+1.7\%$& $+1.7\%$&  $+2.7\%$&$+0.2\%$\\
$d_{78}$&$-1.5\%$& $-1.7\%$&  $-2.6\%$&$+1.2\%$\\
$d_{89}$&$+1.6\%$& $+1.7\%$&  $+1.7\%$&$+2.9\%$\\
$d_{9,10}$&$-0.5\%$& $-0.4\%$ & $+0.2\%$&$+2.2\%$\\
\end{tabular}
\end{table}
\begin{table}
\caption{ Relaxation of Cu(711): experiment and theory}
\label{tab:711rlx}
\begin{tabular}{cccc}
\multicolumn{1}{l}{Relaxation}&
\multicolumn{1}{c}{This work}&
\multicolumn{1}{c}{EAM}&
\multicolumn{1}{c}{LEED \cite{wal01}} \\
\hline
$d_{12}$&$-7.3\%$& $-10.0\%$& $-13\%$\\
$d_{23}$&$-1.5\%$& $-5.3\%$& $-2\%$\\
$d_{34}$&$-14.8\%$& $-9.7\%$& $-10\%$\\
$d_{45}$&$+8.0\%$& $+13.8\%$& $+7\%$\\
$d_{56}$&$-1.0\%$& $-4.5\%$& $-1\%$\\
$d_{67}$&$-1.1\%$& $-4.5\%$&  $-4\%$\\
$d_{78}$&$+1.4\%$& $-4.6\%$&  $+7\%$\\
$d_{89}$&$+1.7\%$& $+8\%$&  $0\%$\\
$d_{9,10}$&$-1.5\%$& $-2\%$&  \\
$d_{10,11}$&$-0.4\%$& $-3\%$&  \\
$d_{11,12}$&$+2.0\%$& $-2\%$&  \\
$d_{12,13}$&$+0.3\%$& $+3\%$&  \\
\end{tabular}
\end{table}

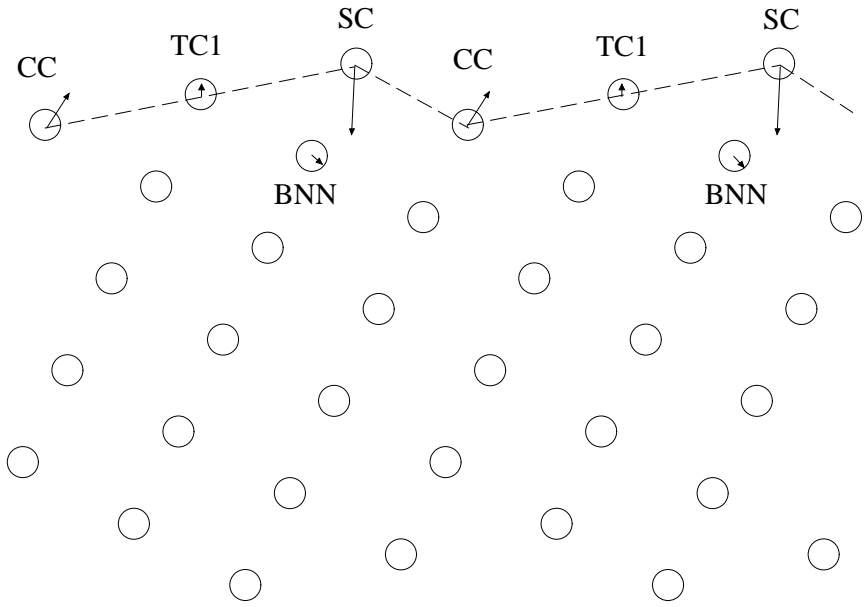
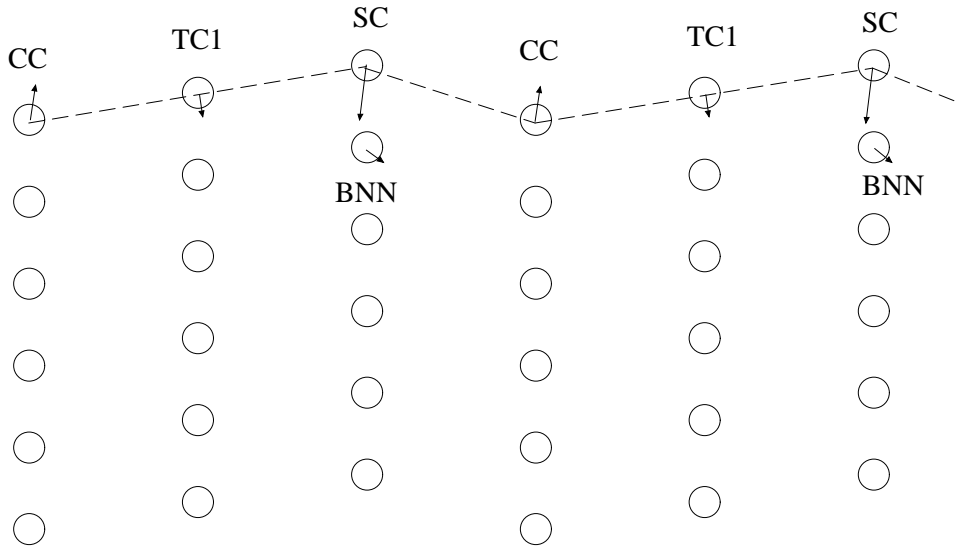
\end{document}